\documentclass[twocolumn,showpacs,prX,nofootinbib,nobibnotes]{revtex4}

\usepackage{graphicx}
\usepackage{dcolumn}
\usepackage{bm}
\usepackage{multirow}

\usepackage{anysize}
\usepackage{graphicx}
\marginsize{1.8 cm}{1.3 cm}{0.5 cm}{3.0 cm}

\usepackage{graphicx}
\usepackage{dcolumn}
\usepackage{bm}
\usepackage{amsmath}
\usepackage{amssymb}
\usepackage{mathrsfs}
\usepackage[latin1]{inputenc}

\usepackage{amsmath}

\newcommand{\dd}{{d}}

\newcommand{\R}{\mathbb{R}}

\begin {document}

\title{Optimal control of evolutionary dynamics}

\author{Raj Chakrabarti}\email{rajchak@princeton.edu}
\affiliation{Department of Chemistry, Princeton University,
Princeton, NJ 08544, USA}

\author{Herschel Rabitz}
\affiliation{Department of Chemistry, Princeton University,
Princeton, NJ 08544, USA}

\author{George L. McLendon}
\affiliation{Department of Chemistry, Princeton University,
Princeton, NJ 08544, USA}


\begin{abstract}

Elucidating the fitness measures optimized during the evolution of complex biological systems is a major
challenge in evolutionary theory. We present experimental evidence and an analytical framework demonstrating
how biochemical networks exploit optimal control strategies in their evolutionary dynamics. Optimal control
theory explains a striking pattern of extremization in the redox potentials of electron transport proteins,
assuming only that their fitness measure is a control objective functional with bounded controls.
\end{abstract}

\pacs{87.23. -n, 87.15.-v, 02.30.Yy}

\maketitle

\section{Introduction}

In a famous paper sent to Charles Darwin and presented before the Linnean Society in 1858 \cite{Wallace1858}, Alfred Russel Wallace -
often considered the co-discoverer of natural selection - proposed that evolution exploits the principles of
feedback control in the generation of biological complexity. Wallace stated:

\bigskip
\bigskip
\textit{"The action of this [evolutionary selection] principle is exactly like that of the centrifugal governor
of the steam engine, which checks and corrects any irregularities almost before they become evident; and
in like manner no unbalanced deficiency in the animal kingdom can ever reach any conspicuous magnitude,
because it would make itself felt at the very first step, by rendering existence
difficult and extinction almost sure soon to follow"}~\cite{Wallace1858}.
\bigskip
\bigskip

During the ensuing development of evolutionary theory, the possibility that nature employs evolutionary control
strategies to maximize the fitness of biological networks has often been discussed in the context of
cybernetics, the study of self-regulation. However, to our knowledge, no direct, quantitative evidence for
Wallace's contention - namely, that evolutionary dynamics itself may be self-regulating - has ever been reported.
In this paper, we provide such evidence, and develop a quantitative physical theory for the interrogation of control
phenomena in the evolution of biological systems.

Evolution is guided by the optimization of fitness measures that balance functionally beneficial properties.
In modern theories of evolutionary dynamics, such as the quasispecies model \cite{Eigen1989} and variants thereof,
the fitness measure of a biological system plays a role analogous to that of the free energy of a mechanical system.
The dynamics of the system, embodied through mutations, seeks to optimize this measure. Recently, with advances in the
understanding of molecular biophysics, increasing attention has been paid to characterizing the fitness measures
underlying the evolution of proteins. For example, simulations of protein sequence evolution have confirmed that
protein cores evolve almost universally to maximize the free energy gap between the folded and denatured states
\cite{Xia2002}. However, for functional properties of proteins and protein networks, the appropriate biological
fitness measures are not so clear \cite{Raj2005}. A current challenge in evolutionary theory is to identify how
the fitness measures of complex biological systems depend on the physical properties of their constituent proteins.

In the hierarchical evolution of protein networks, biological self-organization \cite{Kauffman1989} influences the
dynamics that occur on shorter time scales.  Although most theories of evolutionary dynamics have modeled evolution
as a dynamical system seeking to optimize a potential or free energy, multi-timescale evolution of protein networks may
be modeled within a broader framework as a control problem. Optimal control (OC) theory is generally concerned with the
determination of the time-dependent functional form of the Hamiltonian of a controlled dynamical system that maximizes
a desired objective function \cite{Bryson1975}.  An important difference between a dynamical system and a
control system is that the latter distinguishes between the free dynamics of the system and the dynamics regulated by
controls. In the present case, these controls can take the form of functional protein properties.

The evolution of a biological system may be modeled as a control system if the regulatory functional properties of
its constituent proteins coevolve with the network's overall function. Should the evolutionary dynamics of such a
system demonstrate features indicative of \textit{optimal} controls, this would constitute evidence that the system's
evolution has attained a sophisticated level of self-organization amounting to the solution of an OC problem.  Here,
we show that application of this theory to active site mutations in an enzyme network of central importance for
metabolism - the electron transport chain (ETC) \cite{Anraku1988} - indicates that the redox potentials of electron transport
proteins are controlling the evolutionary dynamics of this network in an optimal fashion, providing insight into the
self-organization of this system.

\section{Artificial evolution of electron transport proteins}

The mitochondrial electron transport chain removes electrons from the high-energy electron donor NADH and
passes them to the electron acceptor $\textmd{O}_2$ through a series of redox reactions involving electron transport
proteins. These reactions are coupled to the generation of a proton concentration gradient across the mitochondrial
inner membrane, which is ultimately used to produce ATP.

Several of these electron transport proteins (for example, NADH-Q reductase, cytochrome reductase and cytochrome c
oxidase) act as both electron carriers and proton pumps, simultaneously catalyzing the transfer of protons against
their concentration gradient. Although the molecular mechanisms of these proton pumps were unclear for some time
\cite{Michel1998,Gennis2004}, recent work has begun to explore these mechanisms, particularly that of the terminal
protein in the chain, cytochrome c oxidase \cite{Belevich2006,Belevich2007}. Belevich et al. \cite{Belevich2007}
studied the proton pump mechanism of this protein in real-time by spectroscopic and electrometric techniques after
laser-activated electron injection into the oxidized enzyme. It was found that the electron transfer reaction to the
primary heme site ("heme a") of this protein raises the pKa of a proton "pump site" amino acid side chain due to the
proximity of the associated negative charge; importantly, the pump protonation site is not in the immediate vicinity
of heme a, and its distance to the heme a plays a role in determining the thermodynamic efficiency of energy transduction. The increased pKa of the pump site draws a proton from the interior of the mitochondrion, with the catalytic assistance of several amino acid residues including Asp124, Glu278, and Lys354 in \textit{P. denitrificans} \cite{Belevich2007}. Protonation of the pump site initiates additional protonation and electron transfer reactions involving distinct sites distal to heme a. In particular, a second protonation induces release of the "pump" proton outside the mitchondrial membrane due to electrostatic repulsion, thus increasing the proton concentration gradient. Note that alteration of the redox potential of heme a would alter not only the thermodynamic efficiency of energy transduction but also the kinetics of its catalysis by Asp124, Glu278 and Lys354, such that mutations around these sites distant from heme a would be required to maximize the efficiency of energy transduction. In extreme cases, mutations that alter the redox potential may even render energy transduction impossible \cite{Puustinen1999}.

Our prior work \cite{Springs2000,Schutt2002} explored the mapping between redox potential and amino acid sequence in the heme microenvironment of ETC proteins.
We pursued a strategy of examining ``evolution in reverse" with the four-helix bundle ETC hemoprotein cytochrome $\textmd{b}_{562}$.  Starting with the evolved protein, variants with replacements at amino acids near the active site heme were created and examined for redox function.  We found two general results. First, within this conserved protein architecture, a range of variation in redox potential $\varepsilon^o$ of about 160 mV could be obtained within two rounds of (reverse) evolution, involving only four residues. Statistical analysis based on Chebyshev's theorem indicates that this range represents, with $> 75\%$ confidence, the total range accessible through mutations at these positions. Second, the wild-type redox potential was not found to be at the middle of the chemically accessible range of reduction potentials \cite{Springs2000,Schutt2002}.  Instead, \textit{wt} $\textmd{b}_{562}$ exhibits a redox potential ($\varepsilon^o = 167$ mV) at the extreme of the chemically accessible range (Fig 1). More generally, artificial mutations on a variety of electron transport proteins of various folds and modes of chemical ligation induce redox potential changes that span ranges between 100-200 mV (Fig 1), typically around 150 mV \cite{Mauk1997,Xiao2000,Chen1999}. Moreover, it is possible to sample the majority of the chemically accessible range through a small number of mutations in the vicinity of the active site, with only minimal concomitant changes to the structure of the scaffold \cite{Schutt2002}.
Most importantly, in nearly every case, these artificial mutations push the redox potential in one direction from the wild-type value (Fig. 1), indicating that this value represents an extremum. In proteins where a few mutations push the potential in the opposite direction (e.g., \textit{Az. Vin.} Ferredoxin and Rubredoxin) it is nonetheless clear that mutation-induced potential changes are strongly biased statistically in one direction from the wild-type potential. Maximum likelihood estimation (MLE) of the underlying redox potential distributions quantifies this conjecture. For instance, in the case of Cyt $\textmd{b}_{562}$, the nonparametric likelihood (see below) that the distribution of redox potentials is unbiased is less than $10^{-8}\%$ (Fig. 1).

\begin{figure*}
\centerline{
\includegraphics[width=5in,height=4.8in]{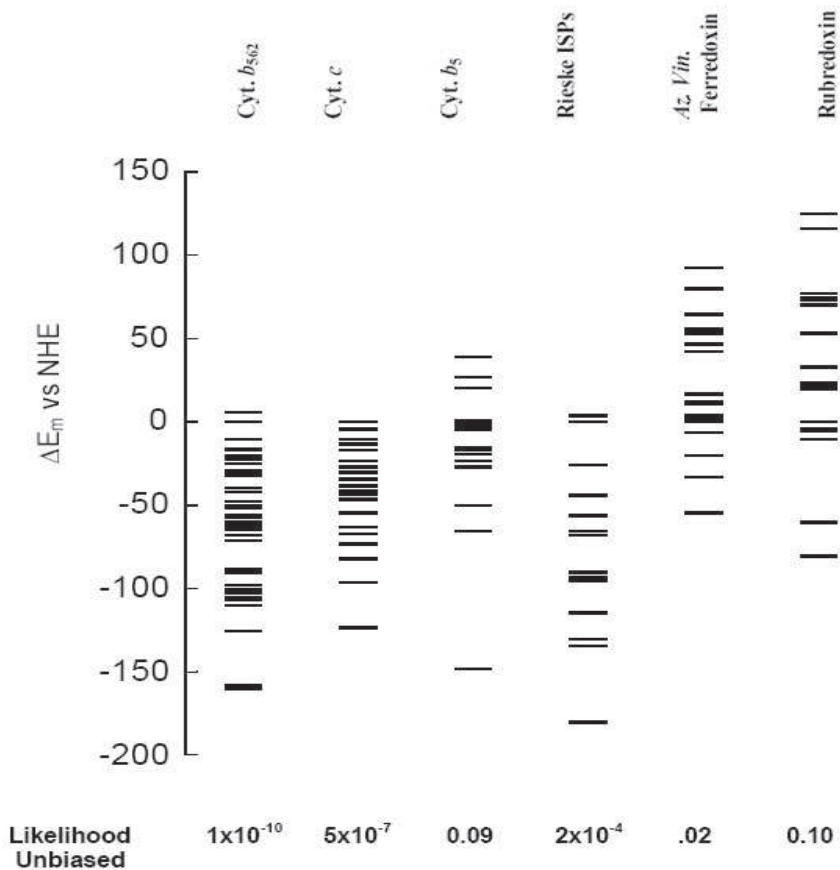}
}\caption{The shift in redox potential from the wild type value ($\varepsilon_{WT} = 0$) for active site mutants of several different cytochromes and iron-sulfur cluster proteins. Maximum likelihood estimation was employed to quantify the extent to which the proteins have evolved toward a redox potential extremum. The likelihood that the true redox potential distribution is unbiased is listed below each protein.}
\end{figure*}

\section{Statistical characterization of mutagenic data}
Redox potential extremization can be quantified by computing the
probability of observing a mutation that shifts the potential to the
opposite side of the putative extremum (wild-type, $\varepsilon_{wt}=0$), under suitable
parametric or nonparametric model distributions. For parametric
models, the likelihood that the underlying probability distribution
of the redox potentials obeyed the model distributions was assessed
by first applying the principles of maximal likelihood estimation
(MLE) \cite{Casella2001} to determine optimal parameters for the
model distributions. The empirical log-likelihood function

$$l(\theta; \varepsilon) = \sum_{i=1}^n \log f(\varepsilon_i\mid \theta) = \sum_{i=1}^n
l(\theta; \varepsilon_i),$$ where $\theta$ is the vector of parameters and $\varepsilon$ is the
vector of redox potential shifts corresponding to $n$ mutations, was maximized by setting the score function vector,
defined as
$$s(\theta; \varepsilon) = \dot l (\theta; \varepsilon) = \sum_{i=1}^n \frac{\partial}{\partial \theta} \log f(\varepsilon_i\mid \theta) = \sum_{i=1}^n
s(\theta; \varepsilon_i)$$ to zero.

The parametric distributions tested included the normal and chi
square probability density functions.
In the case of the normal distribution, analytical expressions exist
for the MLE estimates; the optimal parameter estimates are
equal to the sample mean and sample variance. For the normal
distribution, the student T-test was applied to compute confidence
intervals for the mean $\mu$, and the chi squared test was used to
compute confidence intervals for the variance. For most other
distributions, however, MLE estimates must be determined
numerically. Two different optimization algorithms were used for
determining MLE estimates in these cases: 1) Newton-Raphson (which
employs second derivative information and is the standard MLE
optimization algorithm) and 2) steepest descent / conjugate
gradient. The noncentral chi square distribution probability density
function is
$$f(\varepsilon;\nu; \varepsilon_0) =
\frac{(1/2)^{\nu/2}}{\Gamma(\nu/2)}(\varepsilon-\varepsilon_0)^{\nu/2-1}\exp(-(\varepsilon-\varepsilon_0)/2).$$
The parameters $\nu$ and $\varepsilon_0$ were optimized for the chi square
distribution. Several initial guesses for these parameters were used
to seed the MLE optimization algorithm for the chi square
distribution.

For the optimal normal and chi square distributions, the probability
of observing a mutation that shifts the redox potential to the
under-represented side of the wild-type potential ( $> 0$ for Cyt
b562, Cyt c, Cyt b5, and Rieske ISPs; $< 0$ for \textit{Az. Vin.}
Ferrodoxin and Rubredoxin) was computed by integration of the
probability density in this range. The cumulative distribution
function for the chi square density is

$$F(\varepsilon;\nu; \varepsilon_0) = \frac{\gamma(\nu/2,(\varepsilon-\varepsilon_0)/2)}{\Gamma(\nu/2)}$$

Tables I and II display the mean, variance, optimal MLE
parameter estimates, and probability density below wild-type for
the normal and chi square models. The optimal likelihoods for the normal distributions are generally higher than
those for the chi square model. However, the differences in the chi
square versus normal likelihoods are significantly lower for the
cytochrome proteins. Since the number of parameters in these models
is small, and because of the skewness of the datasets, it is
possible that an asymmetric distribution with properties similar to
the chi square would fit the data better than the normal in these
cases.

\begin{center}
\begin{table*}
\begin{tabular}{c|c|c|c|c|c|c|c|c|c}
\hline
Protein& $\mu$ (mV) & $\sigma$ (mV)  & ML normal & \multicolumn{2}{|c|}{Conf $\mu$ (mV)} &  Conf $\sigma$ (mV)& \multicolumn{3}{|c|}{~~Prob forbid ($\%$)~~}   \\
\cline{5-6} \cline{6-7} \cline{7-10}
 &   & &  & $5\%$   & $10\%$  & $10\%$ & \textmd{opt}  & -$10\% \sigma$ & $10\% \sigma$ \\
\hline
Cyt $b_{562}$&  -60 &   37 &  -186.3 &   $\pm$ 14 & 12 & 7 & 5.3 & 2.2 & 10.7\\
Cyt c&   -49  &   35 &  -104.5 &   19 & 16 & 9 & 8.3 & 3.0 & 13.5\\
Cyt $b_5$&  -18  &   44 &  -78.0 &   28 & 24 & 13 & 34.2 & 27.6 & 37.7\\
Rieske ISPs&  -80  &   48 &  -84.7 &   30 & 26 & 15 & 4.7 & 1.1 & 10.3\\
\textit{Az.Vin.}~Ferrodoxin&  22  &   38 & -96.3  &   22& 19 & 11 & 28.4 & 21.4 & 32.6\\
Rubredoxin&  13  &   24 &  -98.7 &   34 & 29 & 15 & 34.3 & 29.3 & 37.8\\
\hline                                   
\end{tabular}

\caption{\textbf{Statistical properties of mutation-induced redox
potential variations in electron transport proteins: MLE estimation
of normal distribution.} "ML" = maximal likelihood. "Conf" denotes
$+/-$ confidence intervals for mean($\mu$), standard deviation
($\sigma$) calculated by the T-test or chi squared test. "Prob
forbid" denotes cumulative probability density in the "forbidden"
region according to the model ($\varepsilon < 0$ for the cytochromes
and Rieske ISPs, $\varepsilon > 0$ for Ferrodoxin and Rubredoxin).
"Opt" denotes optimal normal distribution; "$10\% \sigma$" denotes
the distributions at the limits of the standard deviation confidence
interval. For example, for Cyt b562 mutants, the 10$\%$ confidence
interval for the mean redox potential of -60 mV is $\pm 12$ mV, and
the 10$\%$ confidence interval for the standard deviation of 37 mV
is $\pm 7$ mV. Under the optimal normal model, the probability of
observing a mutation with redox potential below $\varepsilon=0$ is
$5.3\%$; under the normal model with $\sigma=30$, this probability
is 2.2$\%$, whereas under the normal model with $\sigma=44$, this
probability is 10.7$\%$. The ML of -186.3 indicates that the true
redox potential distribution is less likely to be normal for Cyt
b562 than it is for the other proteins. Typical error bars on redox
potential measurements were $\pm 2-3$ mV.}
\end{table*}
\end{center}

\begin{center}
\begin{table}[h]
\begin{tabular}{c|c|c|c|c}
\hline
Protein & $\nu$  & ~$x_0$~ &  ML $\chi^2$ & ~~Prob $< 0$ ($\%$)~~ \\
\hline
Cyt $b_{562}$& 125.1 &  78   & -233.2 &  0.002 \\
Cyt c&   106.6  &   70 &  -129.8&   0.27\\
Cyt $b_5$& 123.6   &   109&  -101.3&   7.2 \\
Rieske ISPs&  151.7  &  67 &  -125.0 &  0.00005\\
\textit{Az.Vin.}~Ferrodoxin&  121.5  &   119.5 &  -134.3 &   26.7\\
Rubredoxin&  130.0  &   140 &  -198.4 &  22.1 \\
\hline                                   
\end{tabular}
\caption{\textbf{Statistical properties of mutation-induced redox
potential variations in electron transport proteins: MLE estimation
of chi square distribution.} "Prob $<0$" denotes cumulative
probability density in the "forbidden" region according to the
model.}
\end{table}
\end{center}

Figure $2$ displays the optimal parametric (normal and chi-square)
distributions computed by MLE for cytochromes $b_{562}$, c, and
$b_5$, respectively. As can be seen, cytochrome $b_{562}$, which was
most extensively mutagenized (and structurally characterized, below)
displays negligible probability density below the wild type (zero)
redox potential for the chi square model. For the normal model, the
integrated density below zero was computed for the limiting
distributions shifted across the range of means corresponding to a
$5\%$ confidence interval. Even when the variance is upshifted, the
probability of observing a mutation with redox potential below the
mean is $10.7\%$, whereas in the opposite scenario this probability
is only $2.2\%$, providing a clear indication of redox potential
extremization in the natural protein.

\begin{figure}[h]
\centerline{
\includegraphics[width=3.5in,height=3.5in]{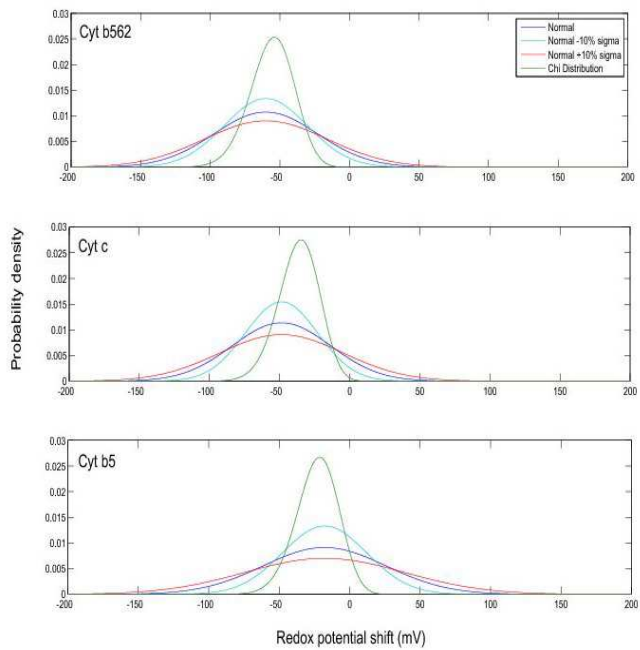}
} \caption{Optimal parametric distributions computed by maximal
likelihood estimation for redox potential shifts. Shifted normal
distributions correspond to upper and lower limits of confidence
intervals for $\sigma$ determined by the chi square variance test.
}\label{cyt562}
\end{figure}

The disadvantage of parametric distributions is that the shape of
the distributions is constrained by the parameterization.
Multinomial distributions were used as canonical nonparametric
distributions. Several nonparametric multinomial distributions were
tested. In this case, probabilities were assigned to discrete redox
potential intervals surrounding the wild-type potential.  The
probability of a mutation falling outside the accessible range of
$~250$ mV was set to zero. The interval width was set to $250/m$ or
$300/m$ mV, where $m$ is the number of intervals. $m$ was varied
between 2 and 8.

The distributions were scored according to the multinomial
distribution probability density function:
$$f(x_1,...,x_m;n,p_1,...,p_m) = \frac{n!}{x_1!\cdots
x_m!}p_1^{x_1}\cdots p_m^{x_m}$$ where $x_i$ denote the number of
observations in interval $i$,  $p_i$ denote the respective
probabilities, and $\sum_{i=1}^m x_i = n$, the total number of
observations.

The likelihoods of several multinomial distributions resulting from
the optimization procedure (for $N=6$) are displayed in Figure 3, for
selected proteins. As can be seen, even for these distributions, where
the probability of observing a redox potential in the "forbidden"
region is generally below $25\%$, the likelihoods of the models are
very low - generally an order magnitude below that of the optimal
model, indicative of extremization of the wild-type potential.

\begin{figure*}
\centerline{
\includegraphics[width=6.5in,height=5in]{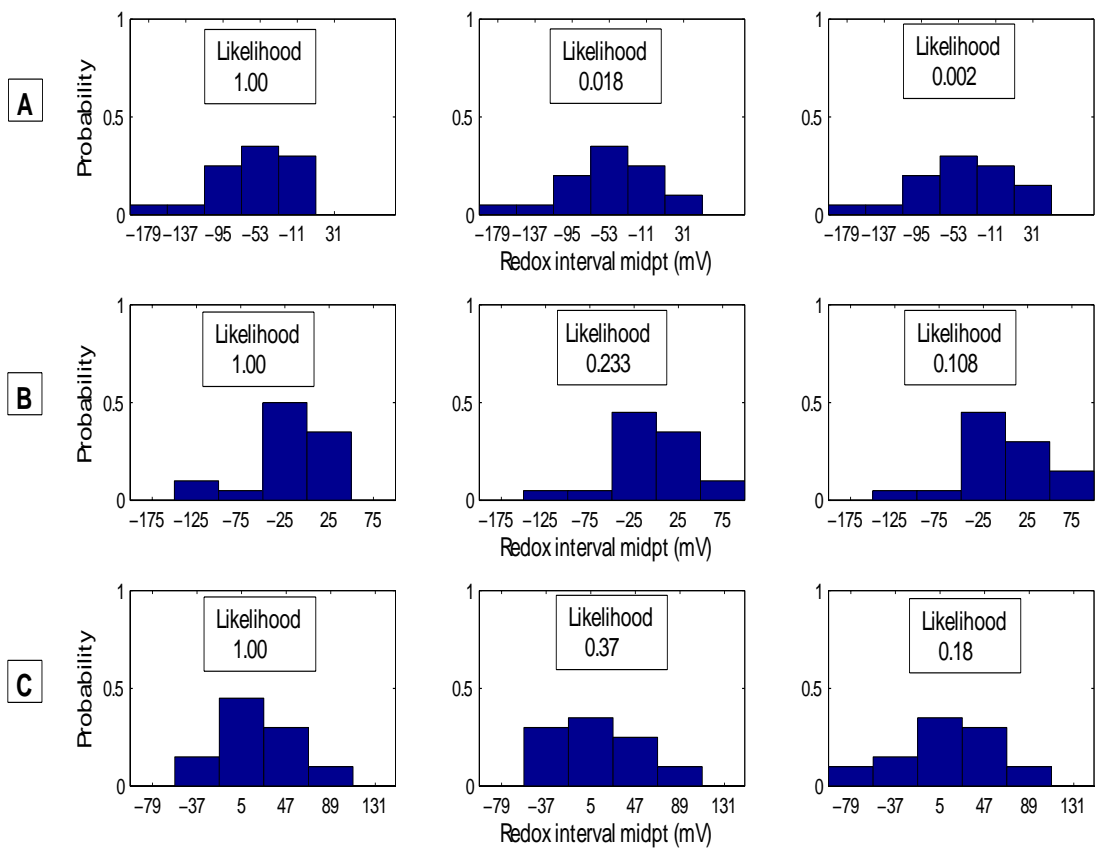}
}\caption{Nonparametric multinomial redox potential distributions
for A) Cyt $b_{562}$, B) Cyt $b_5$, and C) \textit{Az. Vin.}
Ferrodoxin; N=6, with associated relative likelihoods (optimal = 1).
Columns 2 and 3 correspond to "2nd,3rd" distributions from Table
III. Note the rapid falloff in likelihood with increasing
probability density in the "forbidden" region ( $ > 0$ for A,B; $<
0$ for C; redox potentials listed are shifts with respect to
wild-type potential).}
\end{figure*}

\section{Coevolutionary dynamics of redox potential evolution}

The striking observation of redox potential extremization, confirmed by MLE, begs an evolutionary explanation.
It is clear that there is significant evolutionary selection pressure acting on the redox potentials; otherwise,
according to the neutral theory of evolution \cite{Kimura1983}, they would have evolved to maximize robustness to active
site mutations. Wild-type potentials would then lie in the center of the accessible range, and individual mutations
would alter the potentials by only a small fraction of this range - which is not the case.
However, there is no obvious evolutionary advantage to the redox potentials being extremized by this selection
pressure, since maximal fitness (ATP production) follows from maximization of the proton concentration gradient,
which does not bear a simple physical relationship to the redox potentials.

Direct evolutionary selection for extremized redox potentials is implausible
statistically as well as biophysically, based on additional data regarding the distribution of potentials
within the cytochrome c$^\prime$ family, whose members may be perceived as points along a single dynamical
evolutionary trajectory. Two out of four members (Cyt c$^\prime$ \textit{Chr. Vinosum}  and Cyt c$^\prime$
\textit{R. rubrum}) have redox potentials at the lower extreme (-5 and -8 mV, respectively) and two members
(Cyt c$^\prime$ \textit{Alc. Denitrificans} and Cyt c$^\prime$ \textit{Rps. Palustris}) have potentials +100 mV
and +130 mV \cite{Mauk1997}.  Hence, it appears that the redox potentials of naturally-occurring cytochromes are
not only extremized, but may be alternately maximized and minimized during the course of evolution through a process
that requires relatively few mutations. Even if evolutionary selection acted directly on the redox potentials, it
would be necessary to assume the selection pressure oscillates due to environmental dynamics that have no relation
to the known function of the ETC. Such a model is not robust to functional form misspecification of the fitness measure,
and must be rejected if a simpler fitness measure requiring fewer extrinsic parameters can explain the extremization.

Optimal control theory provides an explanation for the observed behavior that is fully consistent
with current evolutionary theory, based on minimal additional assumptions.
In order to apply OC, it is first necessary to formulate the dynamical equations governing the evolution of the ETC.
The terminal oxidation stage of the electron transport chain consists of a linked set of protein-catalyzed substrate
oxidation steps, several of which are coupled to protein-catalyzed proton pump steps. As described above, electron transfer to the redox centers alters the
$\textmd{pK}_a$ of amino acids involved in proton transport and hence \textbf{indirectly} impacts the efficiency
of the proton pumps. The $i$-th enzyme acts on its substrate through a redox process specified by the potential
$\varepsilon_i(t)$ as a function of evolutionary time $t$. The role of this $i$-th enzyme in the fitness measure
can be characterized by its current evolutionary state $x_i$ (i.e., the proton gradient produced by its associated
proton pump) prescribing the functional utility of the enzyme for the energy transduction process.
Since the efficiency of the proton pumps is a function of the redox potentials, it is natural to view the network as
an input-output control system, with the controls consisting of $\vec{\varepsilon}(t) = (\varepsilon_1(t),
\varepsilon_2(t),\cdots,\varepsilon_N(t))$ and the output being the system state vector $\textbf{x}(t)=(x_1(t),x_2(t),\cdots,x_N(t))$.  Evolution is assumed to be maximizing a biologically beneficial function $\Phi(\textbf{x})$ of the
chain's state (i.e., the total amount of ATP produced) both directly with respect to the state $\textbf{x}$
as well as indirectly through the controls $\vec{\varepsilon}(t)$.

	 This evolution of the chain can be modeled in terms of the coevolutionary dynamics \cite{Kamp2002} of
coupled quasispecies sequence families $A$ and $B$, corresponding to each protein's state and control sequences,
respectively. These families are described by the multinomial probability distributions
\begin{eqnarray*}
P_A &=& \{ a_k \mid 1 \leq k \leq n = \kappa^{\nu}\}\\
P_B &=& \{ b_k \mid 1 \leq k \leq m = \kappa^{\mu}\}
\end{eqnarray*}
where $\kappa$ is the monomer alphabet length and $\nu,~\mu$ are the respective sequence lengths.
The probability of producing sequence $A_l$ as an error copy from sequence $A_k$ is given by the elements of the
mutation matrix \cite{Eigen1989},
\begin{equation}
W_{kl} = W_0\left(\frac{w^{-1} - 1}{\kappa - 1} \right)^{d(l,k)},
\end{equation}
$\{k,l \in 1,2,\cdots,\kappa^{\nu}\}$ where $w$ is the fidelity of (base) replication and $W_0 \equiv w^{\nu}$ and
$d(l,k)$ denotes the Hamming distance between sequence $k$ and sequence $l$ (number of monomer positions in which they
differ).

The quasispecies kinetic model assumes sequence growth by first-order autocatalysis and death by first-order decay. We denote by $R_k$ the first-order rate
constant/parameter for autocatalytic amplification, (i.e. replication catalyzed by template $A_k$, of which fraction $W_{kk}$ leads to identical replica)
and by $D_k$ the rate constant for decay of sequence $k$. The quasispecies dynamical equation for the evolution of the probability distribution $P_A(t)$ is then given by
\begin{equation}
\dot a_k = \sum_l W_{kl}R_la_l - D_ka_k.
\end{equation}
In the quasispecies model, the fitness measure $\Phi$ enters implicitly into the evolution equation through its impact on the growth and death
rate constants $R$ and $D$.

In the coevolution of sequences $A$ and $B$, the growth (and death) rates of the DNA sequence encoding the entire
protein that includes subsequences $A_k$ and $B_{k'}$ are explicit functions of only the state
subsequence $A_k$. The probability of a control sequence $B_{k'}$ being replicated (or decaying) is then determined
by the $A_k$ to which it is physically coupled. The probability of subsequences $A_k$ and $B_{k'}$ appearing in
the same strand is $a_k \cdot b_{k'}$. In electron transport proteins containing both a proton pump and a redox
center, these correspond to the pump and active site sequences. We denote the mutation matrix for the second sequence
by $V$, and the growth and death rate constants by $S$ and $E$. Then, in the quasispecies evolution equation for
subsequence $B$, the growth rate constant is completely determined by the matrix $W$, the vector $R$, and the
probabilities $a_k$.

In coevolutionary quasispecies dynamics, the effect of a second coevolving species on the
fitness measure $\Phi$ of the first is typically modeled through a perturbation of the first-order rate constants
$R$, $D$, or both \cite{Kamp2002}. Therefore, in accordance with subsequence $B$ functioning as a control, assume
that sequence $B$ can perturb the fitness measure $\Phi$ such that it affects the growth rate constant of $A$; i.e.,
introduce a $b$-dependence in $R_l$, writing $R_l(b_1,...,b_m)$ or $R_l(P_B)$. We then have:
\begin{eqnarray*}
\dot a_k &=& \sum_{l=1}^n W_{kl}R_l(P_B)a_l - D_ka_k,\\
\dot b_k &=& \sum_{l=1}^m V_{kl}S_l b_l-E_kb_k\\
&=& \sum_{l=1}^m V_{kl} \left[\sum_{q=1}^n \left(\sum_{p=1}^n W_{pq} R_q a_q\right)\right]b_l-\left(\sum_{p=1}^n D_pa_p\right)b_k.
\end{eqnarray*}
We assume the effect is restricted to the growth constant $R$. It is natural to work within a first-order model where $R_l(b_1,...,b_m)$ is a
linear function of the $b_k$'s. In the first-order approximation, $R_l(b_1,...,b_m)$ can be written $R_l^0 + \sum_{r=1}^mR'_{lr}b_r$, such that we get
\begin{equation*}
\dot a_k = \sum_{l=1}^n W_{kl} \left( R_l^0 + \sum_{r=1}^m R'_{lr}b_r\right)a_l  - D_ka_k.
\end{equation*}

Associated with each state sequence $A_k$  is the value $F_k \in \R$ of a component of the associated physical state
vector $\textbf{x}$ of the protein network (respectively $H_k$ for the control vector $\vec{\varepsilon}$). The
expected values of the components of the state and control vectors of the protein network are then
$x_{i}\equiv\langle x_{i}\rangle = \sum_{k=1}^n F_{k}^{(i)} a_{k}^{(i)},\quad
\varepsilon_{i} \equiv\langle\varepsilon_{i}\rangle = \sum_{k=1}^m H_k^{(i)} b_k^{(i)}.$
The tertiary structure of the protein microenvironment surrounding the redox center \cite{Schutt2002} constrains
$\varepsilon_i(t)$ to a finite range
\begin{equation}\label{eqn2}
\varepsilon_i^l(t) \leq \varepsilon_i(t) \leq \varepsilon_i^u(t).
\end{equation}
Because protein tertiary structure is less flexible than secondary structure, it is reasonable to assume the
bounds $\varepsilon_i^l(t)$ and $\varepsilon_i ^u(t)$ vary at a slower rate than the redox potential $\varepsilon_i(t)$
during evolution.

The evolution of the expectation value of the state corresponding to evolution of the distribution
$P_A(t) = \{ a_k(t) \mid 1 \leq k \leq n\}$ is then given by
\begin{eqnarray*}
\frac{\dd \langle x_i(t)\rangle}{\dd t} &=& \sum_{k=1}^n F_k^{(i)} \dot a_k^{(i)}\\
&=& \sum_{k=1}^n F_k \left[\sum_{l=1}^n W_{kl} R_k(b_1,\cdots, b_m)a_k - D_ka_k\right]\\
&=& \sum_{k=1}^n F_k\sum_{l=1}^n W_{kl} R_l(b_1,\cdots, b_m)a_l - \sum_{k=1}^nD_kF_ka_k
\end{eqnarray*}
Now, because the multinomial distribution $P_B$ is sharply peaked with a small variance around a master sequence
$J_{\max}$ (see below), such that $\langle \varepsilon_i \rangle \approx H_{\max} b_{\max}$,
it is reasonable to make the replacement $R_j^0 + \sum_{r=1}^mR'_{jk}b_k \approx R_j^0 + R'_jy_{\max}=R_j^0 + R'_j\varepsilon_a/H_{\max}$.
Furthermore, since the copy fidelity $w \approx 1$, the off-diagonal elements $W_{ij}(~i \neq j)~ \ll W_{ii}$. Under these
approximations, we can write:
\begin{eqnarray*}
\frac{\dd \langle x_i(t) \rangle}{\dd t} &\approx& \sum_{k=1}^n F_k\sum_{l=1}^n W_{kl} \left(R_l^0 + R'_l \varepsilon_a / H_{\max} \right) a_l +\\
&-&\sum_{k=1}^nD_kF_ka_k.
\end{eqnarray*}
If interactions are permitted between state vector components $x_i(t)$, this can be written compactly as:
\begin{equation}\label{eqn1}
\frac{\dd x_i(t)}{\dd t}=f_i(\textbf{x}(t),t)+g_i(\textbf{x}(t),t)\varepsilon_i(t).
\end{equation}

\begin{figure*}
\centerline{
\includegraphics[width=6in,height=4.3in]{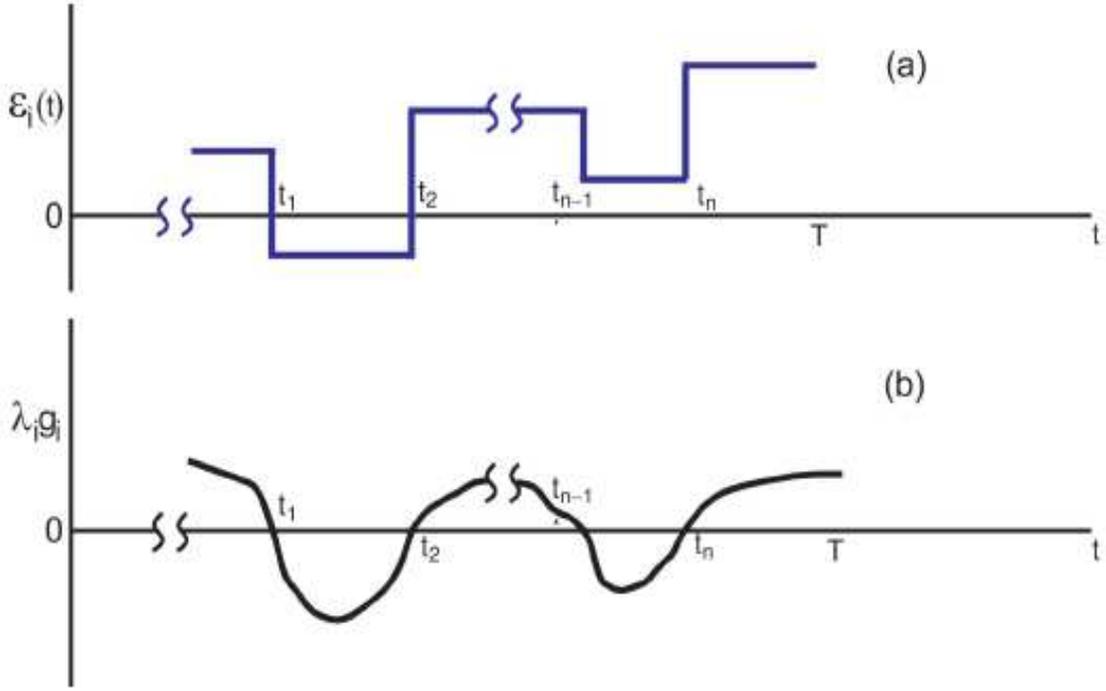}
}\caption{(a) The evolution of the redox potential $\varepsilon_i(t)$ for the $i$-th enzyme within optimal control theory.
The early evolutionary period near $t \approx 0$ is unspecified.  During evolution, the potential ``bangs" from its lower
and upper accessible values, $\varepsilon_i^l(t)$ and $\varepsilon_i^u(t)$, respectively, at critical times $t_1,t_2,\cdots$
where effective mutations have occurred.  The current evolutionary time is $T$.
(b)	The evolutionary time dependence of the product $\lambda_ig_i$ of the Lagrange function $\lambda_i$ and
the control coupling function $g_i$.  The zero crossings of $\lambda_ig_i$ occurs at $t_1,t_2,\cdots$ where
the redox potential $\varepsilon_i(t)$ undergoes evolutionary jumps in (a).
}
\end{figure*}

\bigskip
\bigskip

\section{Optimal control of evolutionary dynamics}

We note that the above evolutionary dynamics framework is based solely on the quasispecies theory and the known function
of the ETC. We now show that the observed extremization implies that the rate parameters $R_{lr}'$ and $D_{ks}'$ have
been set such that the $\vec{\varepsilon}(t)$'s are optimal for maximizing the increase in evolutionary
fitness in a given evolutionary time step $\dd t$. This entails a maximization of $\Phi(\textbf{x})$ with respect to
the controls $\vec{\varepsilon}(t)$, subject to the inequality constraint in Eq. (\ref{eqn2}) and the dynamical
constraint in Eq. (\ref{eqn1}). It is convenient to rewrite the inequality constraint in the form of an equality
through the introduction of so-called slack variables $\xi_i(t)$ where
\begin{eqnarray}\label{eqn3}
G_i(t) &\equiv& G_i(\varepsilon_i,\varepsilon_i^l,\varepsilon_i^u,\xi_i) \\
&=& (\varepsilon_i(t)-\varepsilon_i^l(t))(\varepsilon_i^u(t)-\varepsilon_i(t))-\xi_i^2(t)=0.
\end{eqnarray}
\bigskip
When each of these slack variables $\xi_i(t)$ is allowed to take on arbitrary real values, then the equality constraint in Eq. (\ref{eqn3}) is consistent with Eq. (\ref{eqn2}).  We may now define the fitness measure $J$ as having the
following form:
\begin{widetext}
\begin{equation}\label{eqn4}
J = \Phi(\textbf{x}) + \sum_i\int_0^T\beta_i(t)G_i(t)+\sum_i\int_0^T\lambda_i(t)\left[\frac{\dd}{\dd t}x_i - f_i- g_i\varepsilon_i(t)\right] \dd t.
\end{equation}
\end{widetext}
The introduction of the Lagrange multiplier functions $\lambda_i(t)$ and $\beta_i(t)$ will assure that Eqs. (\ref{eqn1}) and (\ref{eqn3}) are satisfied, respectively.  Equation (\ref{eqn4}) leads to the biological evolutionary process expressed as  $\max_{\vec{\varepsilon}(t)} J$.  Maximization of $J$ can be treated as a problem in the calculus of variations, with the unknown functions being the elements of the vectors $\vec{\varepsilon},\vec{\beta},\vec{\xi},\textbf{x},\vec{\lambda}$.
A variation of $J$ with respect to these functions will produce a set of non-linear equations whose solution would specify the state of the evolving protein network from its initial condition at $t = 0$ to the current time $T$.  Since we have not completely specified the functions $f_i(\textbf{x}(t),t)$ and $g_i(\textbf{x}(t),t)$ in Eq. (\ref{eqn1}), a detailed study of the evolutionary dynamics cannot be carried out here.  However, for our purpose of analyzing the mutation data above, we do not need this level of detail.
It is sufficient to only consider variations of $J$ with respect to $\vec{\xi}$, $\vec{\beta}$, and $\vec{\varepsilon}$, which produce the following equations
\begin{widetext}
\begin{equation}\label{eqn5}
\frac{\delta J}{\delta \xi_i(t)} = -2\beta_i(t)\xi_i(t)=0
\end{equation}
\begin{equation}\label{eqn6}
\frac{\delta J}{\delta \beta_i(t)} = G_i(t)=0.
\end{equation}
\begin{equation}\label{eqn7}
\frac{\delta J}{\delta \varepsilon_i(t)} = \beta_i(t)\left[-2\varepsilon_i(t)+\varepsilon_i^l(t)+\varepsilon_i^u(t)\right]+\lambda_i(t)g_i(t)=0
\end{equation}
\end{widetext}

We may now analyze the evolutionary consequences of these equations.  First, Eq. (\ref{eqn5}) implies that either   $\xi_i(t)=0$ or  $\beta_i(t)=0$.  Considering the first case, $\xi_i(t)=0$, it is evident from Eqs. (\ref{eqn6}) and (\ref{eqn3}) that the redox potential $\varepsilon_i(t)$ must take on the value $\varepsilon_i(t)=\varepsilon_i^u(t)$  or $\varepsilon_i(t)=\varepsilon_i^l(t)$. We may then solve for $\beta_i(t)$  from Eq. (\ref{eqn7}) by first defining $d_i$ as
\begin{eqnarray}\label{eqn8}
d_i&=&-2\varepsilon_i(t)+\varepsilon_i^l(t)+\varepsilon_i^u(t)\\
&=&\left\{
\begin{array}{cl}
\varepsilon_i^u(t)-\varepsilon_i^l(t),&~~ \varepsilon_i(t)=\varepsilon_i^l(t),\\
\varepsilon_i^l(t)-\varepsilon_i^u(t),&~~\varepsilon_i(t)=\varepsilon_i^u(t)
\end{array}\right.
\end{eqnarray}
such that
\begin{equation}
\label{eqn9}
\beta_i(t)=-\lambda_i(t)g_i(t)/d_i(t).
\end{equation}
The second circumstance, $\beta_i(t)=0$, implies that $\xi_i(t)$ is free to take on any value prescribed by Eq. (\ref{eqn3}), given that $\varepsilon_i(t)$ is restricted to the domain specified in Eq. (\ref{eqn2}).  In this case, it is also evident from Eq. (\ref{eqn7}) that $\lambda_i(t)g_i(t)=0$, which is expected to only be valid at discrete times $t=t_n,~n=1,2,\cdots$.  These time points   $t_n$ denote the locations where the control field ``bangs" from one extreme limit of the range to the other in Eq. (\ref{eqn2}) during evolution.

	This behavior may be explicitly seen by considering the curvature
\begin{equation}\label{eqn10}
\frac{\delta^2J}{\delta \xi_i(t)\delta\xi_i(t')}=-2\beta_i(t)\delta(t-t') < 0
\end{equation}
where $\delta(t-t')$ is a Dirac delta function, and the inequality corresponds to requiring that $J$ be maximized. Thus, for the case $\varepsilon_i(t)=\varepsilon_i^u(t)$ in Eq. (\ref{eqn8}), it follows that $d_i < 0$, thereby corresponding to $\lambda_i(t)g_i(t) > 0$, to assure that Eq. (\ref{eqn10}) is satisfied.  Similarly, in the opposite case of  $\varepsilon_i(t)=\varepsilon_i^l(t)$, we have that $d_i > 0$ and that $\lambda_i(t)g_i(t) < 0$.  The points $t_n,~n=1,2,\cdots$  correspond to the times at which $\lambda_i(t)g_i(t)$ changes sign by passing through zero.
This behavior is indicated in Figure 2. The possible evolution of the extremum values $\varepsilon_i^l(t)$ and $\varepsilon_i^u(t)$ is also indicated in the figure.

Importantly, the redox potential data above \cite{Springs2000,Schutt2002,Mauk1997,Xiao2000,Chen1999} are fully consistent with this analysis of bang-bang control behavior.  That is, at the present evolutionary time $T$, each redox potential $\varepsilon_i(T)$ should be at a locally accessible extreme value. The introduction of artificial mutations in the laboratory could then only take $\varepsilon_i(T)$ away from its extreme value in a consistent direction for each protein, as found above.  Moreover, assuming members of the cytochrome c$^\prime$ family lie along the same evolutionary trajectory, their alternatively maximized and minimized redox potentials are consistent with the above model for $t<T$. We emphasize that this finding of optimality is based solely on statistical inference and variational calculus and does not imply anything about the mechanism by which optimality is achieved. 
However, the required tuning of the rate constants $R_{lr}', D_{ks}'$ to optimal constant values is straightforward to achieve via reorganization of the protein's tertiary structure \cite{Schutt2002} through genetic recombination, and avoids the biophysically implausible assumption of direct evolutionary selection for redox potential extremization on an oscillating fitness landscape.


\begin{figure*}
\centerline{
\includegraphics[width=5in,height=4.2in]{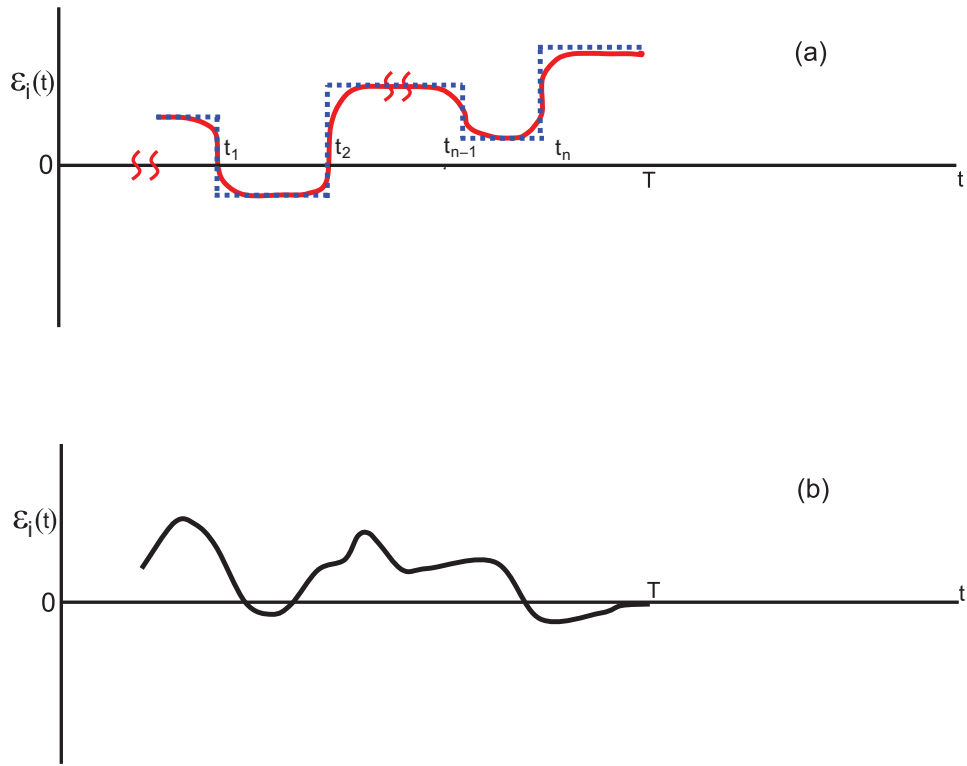}
}\caption{(a) The evolution of the redox potential $\varepsilon_i(t)$ for the i-th enzyme in the case that the evolutionary cost functional is $J' = J-C$, where $J$ is given by Eq. (7) and C by Eq. (8).  The added influence of the cost in Eq. (2) builds in the ability to make smooth evolutionary changes (red curve) in the redox potential while still remaining within its accessible lower and upper values of $\varepsilon_i^l(t)$ and $\varepsilon_i^u(t)$.  The dashed line corresponds to just operating with the cost functional in Eq. (7) (see Figure 2(a) \textbf{FIX}). (b) The cost functional is $J' = J - C$, with C given by Eq. (9). In this case, the genetic pressure on the magnitude of the redox potential permits it to smoothly evolve to have any value within the accessible range between $\varepsilon_i^l(t)$ and $\varepsilon_i^u(t)$.}
\end{figure*}

The OC prediction of bang-bang control behavior is contingent upon the circumstance that the cost functional does not explicitly depend on the controls (except through the Lagrange multiplier that imposes the dynamical constraint), such that
\begin{widetext}
\begin{equation}
J = \Phi(\bf{x}) + \sum_i \int_0^T\beta_i(t)G_i(t) + \sum_i \int_0^T\lambda_i(t)\big[\frac{\dd}{\dd t}x_i - f_i - g_i\varepsilon(t)\big] \dd t.
\end{equation}
\end{widetext}
If auxiliary penalty terms $C$ explicitly depending on the controls $\varepsilon_i(t)$ are introduced, then $J \rightarrow J - C$, and the optimal controls need not be singular, i.e., they may not abruptly "bang" from one extreme to the other.

Two biologically plausible scenarios correspond to:
\begin{equation}
C = \omega \sum_i \int_0^T\big(\frac{\dd \varepsilon_i(t)}{\dd t} \big)^2 \dd t,
\end{equation}
which places a cost on the rate at which control changes occur, and
\begin{equation}
C = \omega' \sum_i \int_0^T \varepsilon_i^2(t) \dd t,
\end{equation}
which places a penalty on the time-average of the control magnitude, in addition to restricting this magnitude to a bounded range.

Figure 3 displays possible optimal controls $\varepsilon(t)$ (redox potentials for the ETC) resulting from these
respective cost functionals. In the case of the former cost, bang-bang control can still be produced, but with a
rounding-off of the sharp corners at the jump times. Biologically, this corresponds to only having an ability to
make evolutionary changes in a gradual fashion, while still taking advantage of the extreme accessible controls as
being biologically most effective.  By contrast, the latter cost may allow the controls
to take on any intermediate values over evolutionary time, since extreme control magnitudes are highly penalized.
The former scenario has a natural structural interpretation, since a given change in a functional protein property,
such enzyme activity, may require multiple (active site) mutations occurring in succession rather than in unison.
The quasispecies error threshold sets a limit on the number of mutations that can be borne by an evolving population
per generation \cite{Stadler1993}. Although bang-bang behavior may not be as apparent in such cases, optimal control
may still be in effect.

\section{Conclusion}

	A natural question concerns the generality of optimal control phenomena in evolutionary dynamics.
Optimal control could in principle be operational in any system where evolution of the central function
of a protein network is coupled to the evolution of an ancillary protein function. Our results indicate
that it is worthwhile to investigate whether the evolutionary dynamics of other biochemical networks with
coupled functions exhibit the characteristic signatures of being under optimal control.  Bang-bang extremization,
while not the only such signature, is simple to detect and provides compelling evidence for underlying OC phenomena.
Such optimal control strategies have a particularly natural interpretation within the general framework of evolutionary
optimization.

The observation that coevolving biopolymer sequences may optimally control each other's evolution raises the
prospect of \textit{artificial} optimal control of evolutionary dynamics. Possible applications include the control of
replication fidelity in nucleic acid amplification reactions and the design of therapeutics that
dynamically regulate the evolution of viral populations.

\section{Acknowledgements}

The authors acknowledge support from the National Science Foundation.


\end{document}